# Facial Biometric System for Recognition using Extended LGHP Algorithm on Raspberry Pi

Soumendu Chakraborty, Satish Kumar Singh, Senior Member, IEEE, Kush Kumar

**Abstract**—In today's world, where the need for security is paramount and biometric access control systems are gaining mass acceptance due to their increased reliability, research in this area is quite relevant. Also with the advent of IOT devices and increased community support for cheap and small computers like Raspberry Pi its convenient than ever to design a complete standalone system for any purpose. This paper proposes a Facial Biometric System built on the client-server paradigm using Raspberry Pi 3 model B running a novel local descriptor based parallel algorithm. This paper also proposes an extended version of Local Gradient Hexa Pattern with improved accuracy. The proposed extended version of LGHP improved performance as shown in performance analysis. Extended LGHP shows improvement over other state-of-the-art descriptors namely LDP, LTrP, MLBP and LVP on the most challenging benchmark facial image databases, i.e. Cropped Extended Yale-B, CMU-PIE, color-FERET, LFW, and Ghallager database. Proposed system is also compared with various patents having similar system design and intent to emphasize the difference and novelty of the system proposed.

## I. Introduction

### A. Motivation

Even with the advent of deep learning in facial biometric recognition, local descriptors still provide a less resource-intensive alternative to face recognition. Local descriptors seem more attractive when recognition is being done in a well-lit constrained environment compared to natural surrounding as they can achieve comparable accuracy with less computational cost. Though looking at the performance of famous descriptors like Local Derivative Pattern (LDP) [1] and Local Gradient Hexa Pattern (LGHP) [2] over world class databases, there is enough room left for improvement. The motivation of this paper comes from the analysis of the evolution of these descriptors. Local Binary Pattern (LBP) [3-5] was proposed to identify the spatial relationship among local neighborhood pixels, Local Derivative Pattern (LDP) extended to derivative space and compared relationship between neighborhood pixels at an angular separation of 45°with a reference pixel. Every other derivative based local descriptor after LDP such as Local Vector Pattern(LVP) [6], Local Tetra Pattern(LTrP) [7] and LGHP added important contributions of their own. Local Gradient Hexa Pattern added discriminating information existing at different radial widths. It combined four first order derivatives at angular separation of 45°with a reference pixel in local neighborhood to get six second order derivatives. Derivative space and an angular separation of 45° between neighborhood pixels viz. 0°, 45°, 90°, 135° is the common element among several recent derivative-based facial descriptors. In this paper, we propose angular separation of 90° among local neighborhood pixels in first order derivative space for calculating second order derivative. Increased angular separation in comparison to LGHP results in enhanced discriminating power. The descriptor thus obtained shows improvement over LGHP on several world class databases.

Execution time of Local descriptors depends on the size of an image and the number of images it is compared against in the database. LGHP introduced the concept of including information available at higher radial distances. Although this novel aspect lead to better performance compared to other descriptors but resulted in much higher execution time. The paper proposed a serial model which doesn't take advantage of several independent steps in the whole face recognition process. This paper explores the larger local neighborhood as compared to LGHP and proposes a new descriptor with improved accuracy and execution time when it is executed in parallel with multiple threads working on separate tasks compared to serial execution. Here we consider thread division between the two major tasks, feature extraction and classification. The proposed parallel execution model can also be applied to similar descriptors like LGHP, LDP, LVP and others.

Another part of this paper deals with designing a facial biometric system based on client-server architecture. The motivation of this comes from the recent introduction of community supported devices like Raspberry Pi [8] which provides an easy Do-It-Yourself method to test various algorithms in real world scenario and also, in turn, show how easily a biometric system can be made. Although face recognition using Raspberry Pi or a portable device has been proposed before [9], but here the novelty lies in a scalable implementation using client-server model with no local face databases stored in the device or sent across the network for added security. To highlight the uniqueness of the proposed system, it is compared with existing patents in similar area.

The motivation behind proposing this system can be

Authors are affiliated to Indian Institute of Information Technology, Allahabad-211012, India(e-mail:kush.kr889@gmail.com, sk.singh@iiita.ac.in, soum.uit@gmail.com)

summarized as
1. The existing LGHP descriptor, even though performed better than most of the existing descriptors, captured redundant features. This redundancy hampered the accuracy.
2. The LGHP algorithm is serially executable algorithm, which doesn't take advantage of several independent steps in the entire face recognition process.
3. This paper explores the larger local neighborhood as compared to LGHP and proposes a new descriptor with improved accuracy and execution time when it is executed in parallel with multiple threads working on separate tasks compared to serial execution.
4. Although face recognition using Raspberry Pi or a portable device has been proposed before [9], but in the proposed system novelty lies in a scalable implementation using client-server model with no local face databases stored in the device or sent across the network for added security.

*B. Related Work*

Face Biometric System being a patentable invention, there are several existing patents in this area. Real-time facial recognition and verification system [10] defines a system with no mention of remote storage of image features. This patent keeps body tone of a number of people and then tries to find a region of interest in the captured image. The method says nothing about how images are captured, stored or transferred. Another similar patent, Pose-invariant face recognition system and process [11] deals with extracting face regions and verification using a neural network and mentions the use of a distributed system but no specifics about data exchanged between clients and the server system is provided. It simply states that program modules may be located in remote servers. Face recognition system [12] proposes a standalone system and mentions no remote system or servers for processing data. The method proposed in this patent works on an image with a large number of people. In the first phase, the method detects if the image captures any person at all and then finds if that person can be identified by comparison with images of the individuals stored in the database. Another patent titled Face recognition system and method [13] is about generating a 2D image from three-dimensional model and provides no information about scalability or security. A patent [14] with client-server paradigm like the one proposed in this paper talks about building a system capable of pattern matching between a less noisy face image and a noisier face image obtained in different imaging conditions. Boaz J. Super et. al. [15] proposed a method and apparatus for selective transmission of color information to a remote recognition system. The patent talks about remote recognition but proposes to send images across the network and not features.

Local Binary Pattern (LBP) [3-5] originally proposed for texture classification was the first to be used as local face descriptor for its discriminative power and computational simplicity. It introduced the concept of local spatial pattern. Local derivative Pattern introduced the concept of high order derivatives to encode directional features based on local derivative variations. Local Vector Pattern (LVP) [3] used the concept of higher order derivative in the same four directions as introduced by LDP and encoded pair wise vectors using comparative space transform. Local Tetra Patterns (LTrP) [4] proposed has similar concepts to that of LDP. Local Gradient Hexa Pattern (LGHP) [2] introduced a concept of a discriminative relationship between a reference pixel and its eight neighborhood pixels. All the descriptors after LBP experimented with different concepts revolving around derivative of a facial image in the same four directions, viz. 0°, 45°, 90°, 135°.

*C. Major Contribution*

LGHP is a breakthrough local face descriptor that shows improvement in accuracy on benchmark facial image databases compared to its predecessors like LDP, LVP and LTrP by exploiting the discriminating information between the different derivatives at 0°, 45°, 90° and 135°. Major contribution of this paper is to enhance the performance by modifying the binary structure of LGHP in the local neighborhood of a facial image. The entire process of image recognition through any descriptor like LGHP is inherently independent at several levels hence different threads can be defined at different radial distances from a pixel, feature extraction level and classification level. It's apparent from result analysis that the improvement achieved at a radius greater than 3 is not significant enough to justify the increase in radial width, the upper bound on the radius of the proposed descriptor is 3. Thus, we are limited to maximum of three threads if we parallelize with different threads defined at different radial widths. Hence, this paper explores distributing threads between feature extraction and classification. Another contribution of this paper is the execution time reduction that can be achieved using proposed methodology. This paper also proposes a client-server based biometric system implementation which keeps no local database of images and never sends them across network to the central server for verification. The proposed system requires no downtime of central server for adding a new client system responsible for face recognition. During registration process, facial images are exchanged on the network between client and server but the images are registered only when the user id sent by the server after successful registration matches the one generated by the client system. These aspects in our system set it apart from other proposed facial biometric systems.

## II. PROPOSED SYSTEM

The system proposed in this paper is built on client-server architecture. The client system consists of Raspberry Pi 3 model B running Raspbian OS and server is a Lenovo all in one desktop running Intel(R) Core TM i3 3320 CPU with 6GB DDR3 RAM 1TB HDD with Ubuntu 16.04 as the operating system. Raspberry Pi are single board computers made by Raspberry Pi foundation. The proposed system uses their most recent and advanced version as of May 2017, Raspberry Pi 3 model B. This model has a 1.2 GHz 64/32-bit quad-core ARM Cortex-A53 processor, 1 GB LPDDR2 RAM at 900 MHz and Broadcom Video Core IV graphics processor. The model also has Ethernet, Bluetooth, Wi-Fi, 4 USB ports, audio output, mini HDMI, GPIO pins, Display Serial Interface and MIPI CSI-2 camera interface. Pi 3 mode B has a micro USB input for charging via its 2.5-amp AC adapter. In conjunction with Raspberry Pi 3 model B, the system uses a 7-inch official

### A. Interconnection

As shown in Fig. 1, various components of the proposed system consist of Raspberry Pi 3 model B, touch screen display, Pi camera, micro USB cable, power adapter and a micro-SD card. At first, Raspberry pi, henceforth to be named as just Pi or RPI or the client system, is mounted on the touch screen. It's then connected to the display using a micro USB cable going from its micro USB port to USB port on the display. We do this to power Pi using the display. Next, we connect the ribbon cable attached to the display to the DSI port on the pi. The Pi camera is then connected through its Raspberry Pi touch screen, Raspberry Pi camera module having Sony IMX219 8-megapixel sensor and a 2.5-amp AC adapter.

ribbon cable to CSI port. NOOBS OS installer is downloaded from raspberrypi.org on a micro-SD card and inserted into the micro-SD card slot on the board. Finally, we connect the power adapter to the micro USB port of the display, which automatically enables the device. Fig. 2 shows the device after all interconnections were completed.

We then proceed to install Raspbian OS, OpenCV, and QT on Raspberry Pi. QT is a cross-platform application framework which is used to develop the user interface in C++. OpenCV is a computer vision library used to read images and extract pixel intensity information.

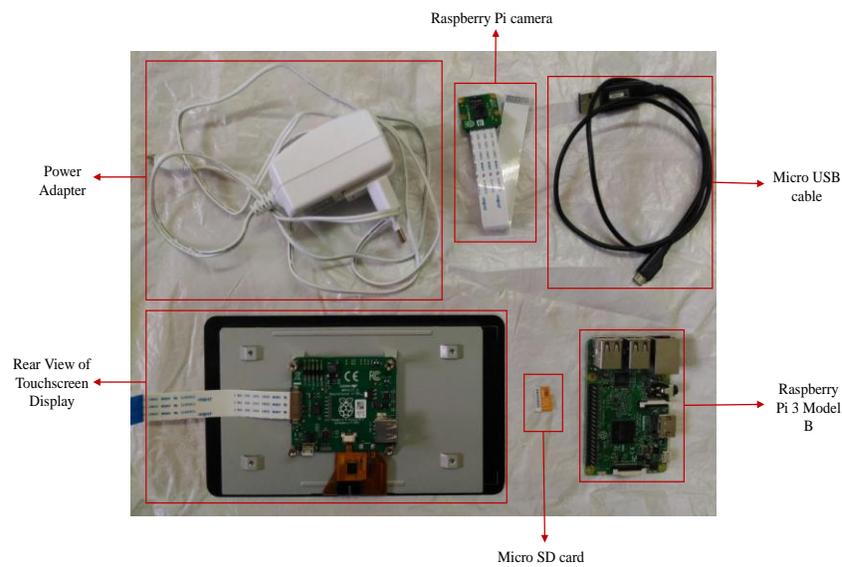

Fig. 1. Various components of the proposed system.

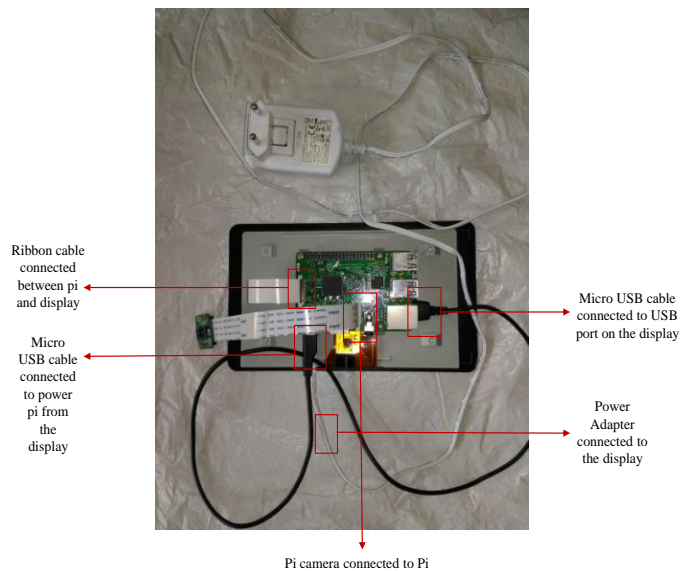

Fig. 2. Assembled biometric System.

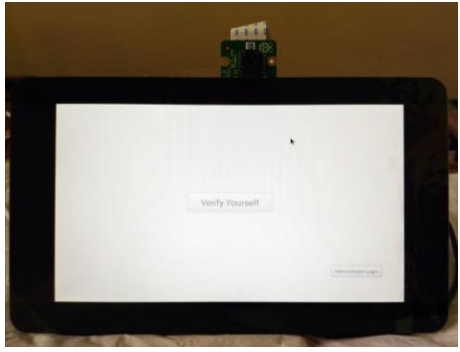

Fig. 3. Face recognition application running on the system.

*B. Design Principles*

The application runs on the assembled hardware as shown in Fig. 3. The proposed system is based on client-server architecture. Fig. 4. shows the hardware architecture of the proposed system. The user image database is located at the server, the client system never stores any image locally. No user image is sent over the network during the verification process. Verification process instead sends image features over the network to the server. The server program as illustrated by the algorithm in Fig. 7 contains logic to interact with the client via TCP, extract features from database image using ELGHP and compare it against the feature sent over by the client system for user verification. The client system has logic to capture user image, extract image features and send over to the server if it's meant for verification or send the raw images if a new user is being registered. Fig. 7 shows how extracted features of all the images of the database are matched with the feature of the queried image at the server. Line number 6 in the pseudo code extracts feature of the queried image. Line number 9 to 11 establishes the connection with the server to transfer the extracted feature to the server. In line 12, server receives the feature and saves in a vector. The distances of these features from the feature of the queried image are computed in lines 13 to 20. The name of the minimum distance image is extracted in lines 21 to 22. This extracted name is matched with the given user-id in line 26 for authentication. Fig. 6 shows the client side logic via pseudo code. The pseudo code in Fig. 6 is pretty straight forward. The feature of the queried image is extracted at line 9 and transferred to a server site function called FaceRecServer (shown in Fig. 7) at Line 11. The client system allocates a unique user id to each user at the time of registration and expects the same at the time of verification along with facial data. This is done to introduce an added level of security. Both client and server side contain the implementation of ELGHP algorithm elaborated in section-III, for feature extraction of user image and database image respectively. More client systems can also be added without interfering with the existing client-server system setup. The novelty of the system in comparison to patents described in the previous section is the transmission of image features instead of raw images for verification with remote authentication during user verification and two-factor authentication with user-ID and facial data. Fig. 5 illustrates the flow-chart of client-side behavior of the proposed system.

## III. PROPOSED EXTENDED LGHP

Extended LGHP or ELGHP is a variant of LGHP [2] which works on neighborhood pixels at an angular separation of 90° from the reference pixel. Four first order derivatives are computed at an angle $\alpha = \delta \times 0°$, $\delta \times 45°$, $\delta \times 90°$ and $\delta \times 135°$ of a reference pixel. Eight neighbors at a radial distance of $\delta = 1$ or 2 are then compared across different derivative space. $\delta = 1$ is the standard LGHP and ELGHP is computed at $\delta = 2$.

Second order ELGHP is then calculated by making a pair wise comparison of these four derivatives resulting in six total binary vectors corresponding to each pairs. Pair wise comparison is done using a simple encoding function defined as

$$F\left(G_{\alpha,d}(P), G_{\beta,d}(P)\right) = \begin{cases} 1, & if\ G_{\alpha,d}(P) > G_{\beta,d}(P) \\ 0, & else \end{cases} \quad (1)$$

Where $G_{\alpha,d}(P)$ is the derivative at an angle $\alpha$ and distance $d$ for the reference pixel $P$. Similarly, $G_{\beta,d}(P)$ is the derivative at an angle $\beta$ and distance $d$ for the reference pixel $P$. The encoding function F takes as input, two first order derivatives $G_{\alpha,d}(P)$, and $G_{\beta,d}(P)$ of 9 pixels (8 neighborhood pixels and reference pixel) calculated at different angles and the same distance d from the reference pixel P and generates a string of 1/0 after comparing each value. The result obtained is a 9-bit binary value which is then converted to decimal. Spatial histogram of these six matrices corresponding to each pair of four derivatives is then obtained after keeping the decimal value of the 9-bit binary number at pixel locations. The histograms obtained are concatenated to generate a vector of length 512×6 since we have 9-bit values in each of the six matrices. The process continues for radial distance d = 2 and d = 3. Finally, all d value vectors are concatenated resulting in a single vector of length 512×6×3 which we refer as the feature vector of an image. Similarity measure which is defined as

$$D_{L1}(v1, v2) = \sum_{i=1}^{p} |v1_i - v2_i| \quad (2)$$

Here $D_{L1}$ is the L1 distance computed on two vectors v1 and v2. 1NN is the used to calculate the L1 distance between test image and DB image. If the class of the image to which the test image is found closest among all database images is same as that of the test image, we call it correct match else mismatch. Fig. 8 shows the above calculation as an example for a reference pixel.

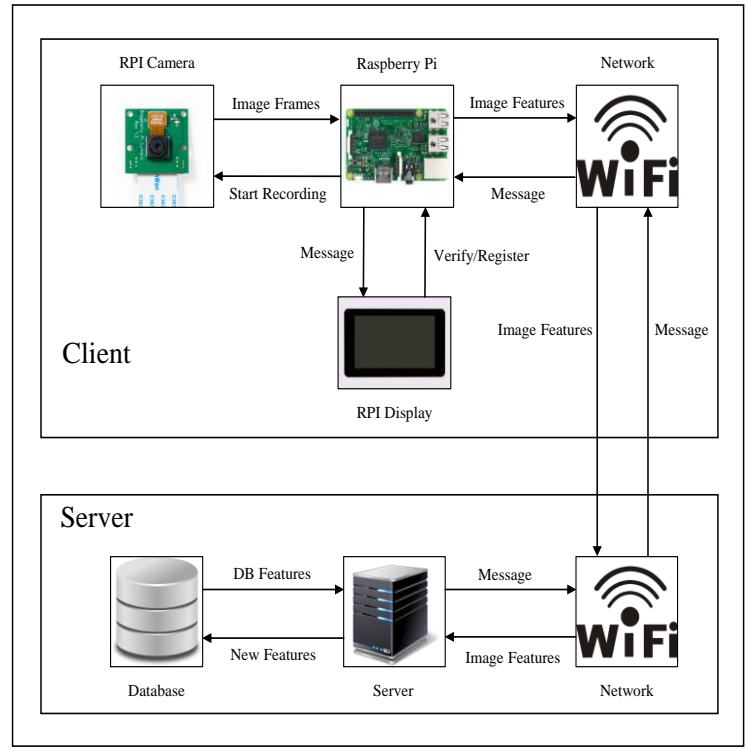

Fig. 4. Hardware Architecture of the proposed facial biometric system. The diagram contains the information flow between various components of client and server system for user registration and verification.

## IV. IMPLEMENTATION LEVEL PARALLELISM

ELGHP has several inherent properties that give various opportunities to parallelize the whole process of feature extraction and classification. Three major computations that are done in ELGHP are second order derivative calculation at each pixel of an image, second order derivative calculation at different radial distances at each pixel and image classification.

Following are the ways thread allocation can be done:

1. Different threads can work at different radial distances from a pixel, one image at a time. We can have another thread working to concatenate the values achieved. Classification can proceed subsequently.

2. The image can be divided into as many parts as threads available with each thread working to extract features from the pixels in its part of the image, one image at a time. Classification can proceed subsequently.

3. Threads can be equally divided into the major tasks, feature extraction and image classification. The total number of image for feature extraction and classification is then divided equally among their respective threads groups.

We now propose the design of the 3rd option to parallelize ELGHP. ELGPH implementation was done using C++11, OpenCV and compiled in G++ 5.4.0 in Ubuntu 16.04. In order to better compare multi-thread vs single thread performance, both the implementations were done as close as possible. As we have four threads in our test machine, tests were performed with thread count = 1, 2 and 4. When thread count = 1, it is considered as serial execution. As illustrated in Fig. 9, all the implementations with thread count greater than one had threads divided equally between feature extraction and classification. C++ std::thread class was used to perform multi-threading and one-dimensional std::vector associate container of C++ was used to hold features of an image. We had an array of std::queue container defined with size equal to (no. of thread)/2. The size of the queue container array was kept such to ensure that each thread working to extract and push features have their own respective queue to avoid any write contention. The queue stores a pair of values, image name, and image features. Half of the threads allocated for feature extraction continue operation pushing image features of their share of images in their respective queue, whereas the other half of total threads iterate trying to find if there is any new image feature present in any of the queue, if it is, it's extracted out for 1NN classification against features of DB images. Terminating condition for the first half of the threads is when there is no image left to extract and for the other half is when the count of images extracted from all the queues is equal to the total number of images.

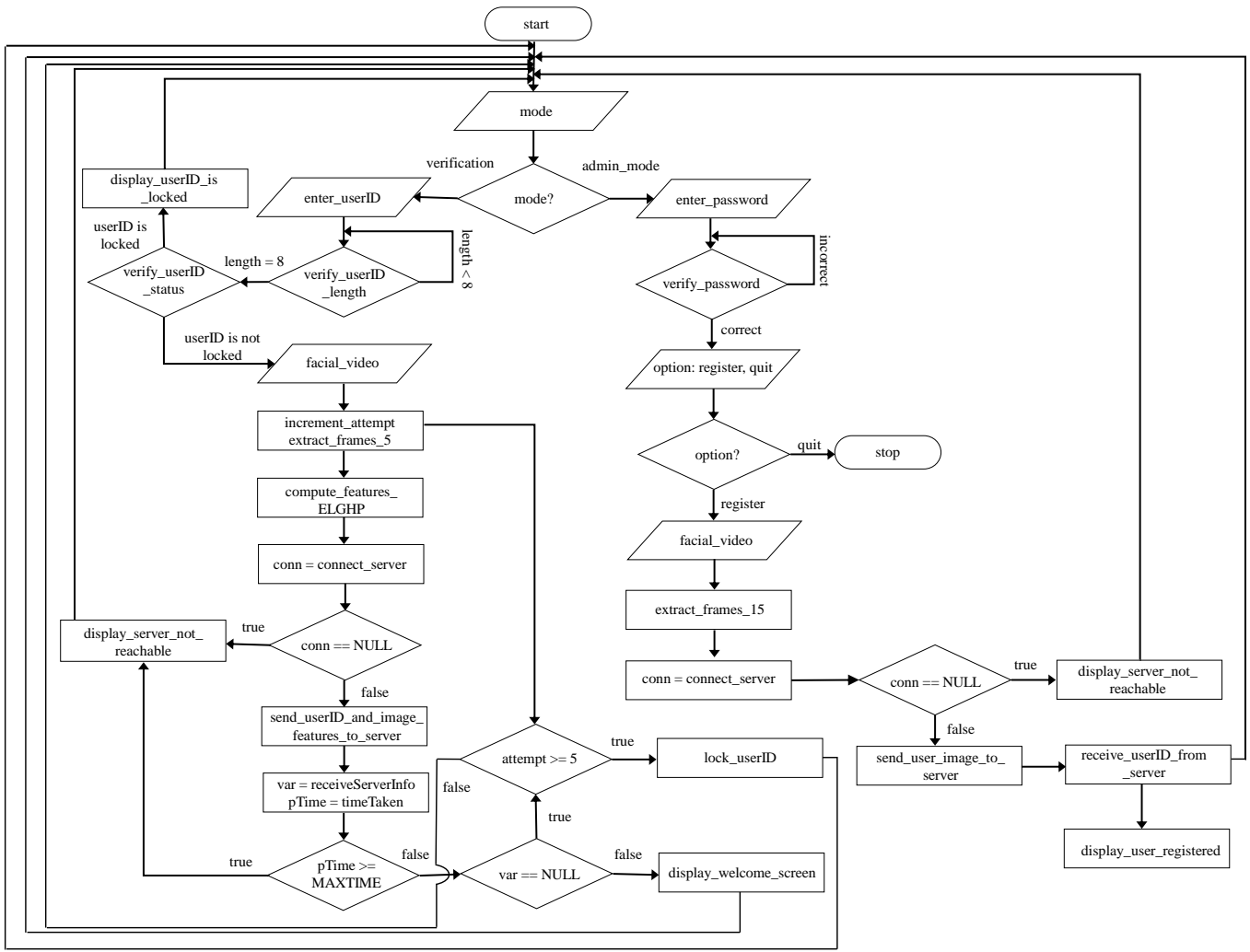

Fig. 5. Flow-chart of client module of the proposed system

Algorithm 3. FaceRecClient

1. Input:
2. Output: matched
3. Intial:   imgFeatures ← [][], matched ← false, imgFrames ← []
4.
5. Begin
6.    userId ← receiveUserIdFromUI()
7.    facialVideo ← recordVideo()
8.    imgFrames ← ExtractAndSelectFrames(facialVideo)
9.    imgFeatures ← ELGHP(F, 0, TESTIMGCOUNT, NULL)
              // // serial execution of ELGHP
10.   FaceRecServer(userId)        // via TCP/IP connection
11.   FaceRecServer(imgFeatures)    // via TCP/IP connection
12.   matched = receiveFromServer()   // via TCP/IP connection
13.   return matched
14. end

Fig. 6. Pseudo-code of user verification process in client module

```
Algorithm 2. FaceRecServer
  1.  Input:    db Image F[], db Image Name imgName[]
  2.  Output:   matched
  3.  Intial:   dbImgFeature ← [][], DIST ← 3, HEXA ← 6,
               FEATURECOUNT ← 512, matched ← false, userImgCount ← 5
  4.
  5.  Begin
  6.    dbImgFeature = ELGHP(F, 0, DBIMGCOUNT, NULL)  // serial execution of ELGHP
  7.    for m ← 1 : ∞
  8.       matched ← false
  9.       acceptClientConnection()                                  // via TCP/IP connection
 10.       makeThread()                                  // spawn a new thread to server this client
 11.       userid ← receiveClientData()                              // via TCP/IP connection
 12.       userImgFeatures[] ← receiveClientData()                   // via TCP/IP connection
 13.       for i ← 1 : userImgCount                                  // 1NN
 14.          minDist ← infinity
 15.          for j ← 1 : size(dbImgFeature)
 16.             tempMin ← 0
 17.             for k ← 1 : FEATURECOUNT×DIST×HEXA
 18.                tempMin = tempMin + |userImgFeatures[i×FEATURECOUNT×DIST×HEXA + k] -
                        dbImgFeature[j][k]|
 19.             end
 20.             if tempMin < minDist
 21.                match = imgName[j]
 22.                minDist = tempMin
 23.             end
 24.          end
 25.       end
 26.       if userid == match
 27.          matched = true;
 28.       end
 29.       return matched
 30.    end
 31.  end
```

Fig. 7. Pseudo-code of user verification process in server module

| 2 | 5 | 3 | 9 | 5 | 4 | 3 |
| 1 | 3 | 8 | 7 | 6 | 2 | 4 |
| 5 | 4 | 1 | 6 | 9 | 2 | 3 |
| 4 | 9 | 8 | 2 | 7 | 5 | 3 |
| 4 | 2 | 3 | 6 | 5 | 6 | 9 |
| 6 | 1 | 7 | 4 | 5 | 1 | 8 |
| 5 | 3 | 9 | 8 | 5 | 6 | 7 |

| -5 | 2 | 7 | -3 | -5 | 6 | -3 | 1 | -1 |
$G^1_{0°,1}(P)$

| -4 | -2 | 3 | -1 | -7 | 7 | -5 | 4 | -2 |
$G^1_{90°,1}(P)$

| -6 | 5 | 3 | 5 | -3 | -1 | 1 | 3 | -1 |
$G^1_{180°,1}(P)$

| -4 | 2 | 2 | 4 | -7 | 5 | -4 | 2 | 0 |
$G^1_{270°,1}(P)$

(a) (b)

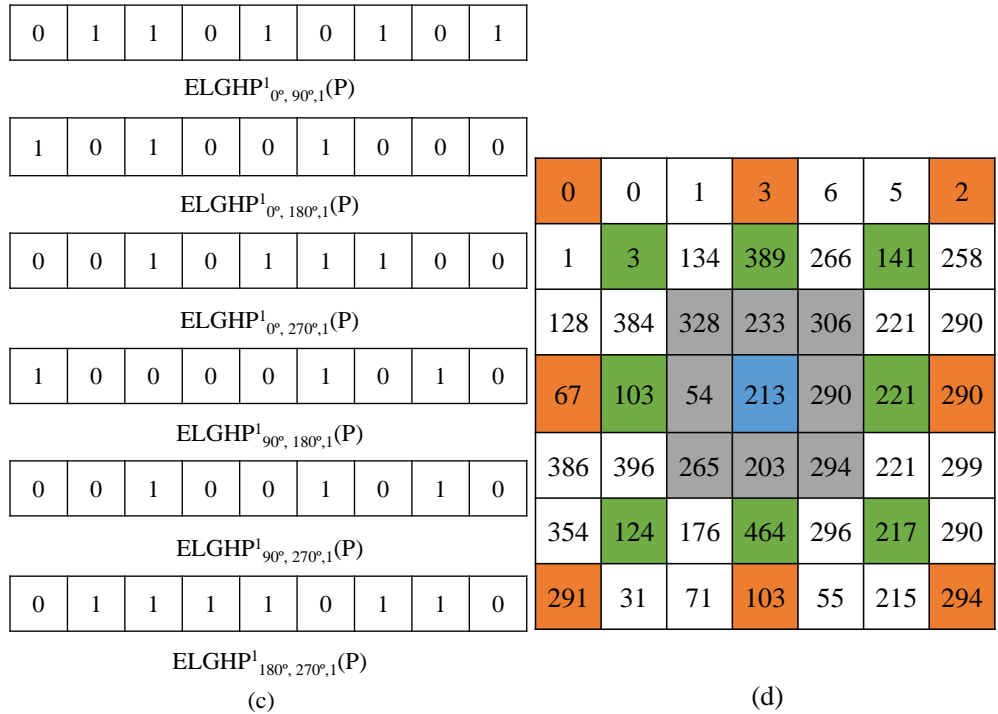

Fig. 8. (a) Sample image, neighbors at radial distance d = 1 shown in gray, d = 2 shown in green and d = 3 shown in orange from a reference pixel colored in blue, (b) First derivative at different angular separations of the reference pixel, (c) Six pair wise combination of four first order derivative encoded by the function in eq. no. (1) (d) Transformed image with pixel intensity values replaced with the corresponding $ELGHP^1_{0°, 90°, 1}(P)$ values.

As feature extraction speed is partially dependent on secondary storage speed whereas image classification is independent of it, efficient thread distribution between extraction and classification can also be other than what we found in our test set up. This distribution is dependent on the speed difference between primary and secondary storage, the dimension of test images and the number of DB images. Secondary storage speed and dimension of test images impact time taken for test feature extraction whereas the number of DB images only effects classification speed. Primary storage speed effects both classification and feature extraction. The overall novelty of the system, descriptor and implementation proposed can thus be summarized as:

 1) The proposed facial biometric system transfers features of an image calculated using ELGHP to server for verification instead of user image. The existing systems either transmit image over the network [15] or locally handle the verification process [9]. The proposed method reduces dimension of the transmitted data and increases security, as network spoofing for user data cannot compromise user privacy.

 2) The system is both scalable and secure. Any added client system only requires the know-how to interact with the server. The server which spawns individual threads to serve every unique client thus requires no downtime.

 3) The descriptor proposed is a unique variant of LGHP with increased discriminating power. The proposed descriptor shows significant and consistent performance improvement on several world class databases.

 4) Implementation level parallelization explores unique approach to reduce execution time of the descriptor proposed. This approach can also be applied on similar descriptors.

## V. PERFORMANCE ANALYSIS

### A. Recognition Accuracy

The performance of ELGHP has been analyzed by finding average recognition rate on various world class databases. Several different disjoint test and database splits were considered. To compute the recognition accuracy the database has been divided into test set and gallery set. Test set is called the Test and gallery set is called the DB. 20-80, means 20% images are in the test set and 80% images are in gallery set. Recognition accuracy is defined as

$$Rr = (N_i/T_i) \times 100 \quad (3)$$

Here Rr is the recognition accuracy, $N_i$ is the number of images correctly matched and $T_i$ is the total number of test images. All the facial images were resized to 64×64. Histograms of 512 bins are concatenated to generate the descriptors. Experimental results were computed on a system running Intel(R) Core TM i3 3320 CPU with 6GB DDR3 RAM 1TB HDD with Ubuntu 16.04 as the operating system. The algorithm was coded in C++ with OpenCV. The various databases used for the analysis were LFW [16], FERET [17-18], CMU-PIE [19], Cropped Extended Yale B [20-21] and Gallagher [22]. The average accuracy was computed by splitting the database 10 times into different test and gallery sets. For instance the database was split into test set (of size 20%) and gallery set (of size 80%) 10 times and each time recognition accuracy was computed and then the average of all these 10 accuracies has been reported in the figures. All the accuracies reported were computed at radius = 3 for both LGHP and ELGHP.

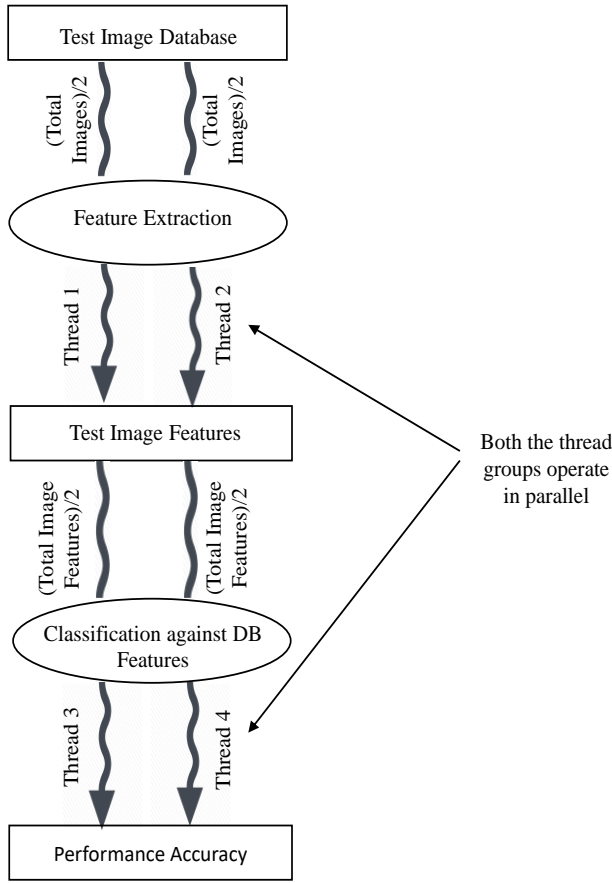

Fig. 9. Thread allocation in parallel implementation of ELGHP.

### A1. Recognition Accuracy on LFW database

LFW or Labeled Faces in the Wild Home database is a database meant for studying the problem of unconstrained face recognition. Images of individuals having at least 20 images belonging to it were taken resulting in a total of 3023 images. The images present in this database has been taken in natural surrounding and frequently contains more than one face in a single image, this adds another challenge to already existing issue of natural surroundings. The result of these challenges leads to a low accuracy of all well-known descriptors on LFW. As shown in Fig. 10, ELGHP working on 20-80 test database split, with 602 images in test and 2421 images in the database gives 15.4% accuracy with R = 3 and 20% test size compared to LGHP which gives 14.5% accuracy.

### A3. Recognition Accuracy on CMU-PIE database

CMU Pose, Illumination, and Expression (PIE) Database contains 41,368 images of 68 people, each person under 13 different poses, 43 different illumination conditions and with four different expressions. We measure accuracy separately on sets containing different expression, lights, and illumination. ELGHP has 96.7% accuracy on CMU-PIE-Expression with R = 3 and 20% test size compared to LGHP which has 95% accuracy on the same. As shown in Fig. 12, improved accuracy trend continues to both CMU-PIE-Lights on which ELGHP has 97.1% accuracy with R = 3 and 20% test size versus LGHP which has 96.3% accuracy, and CMPU-PIE-Illumination on which ELGHP records higher accuracy of 59.1% with R = 3 and 20% test size compared to LGHP which has the accuracy of 58.2%.

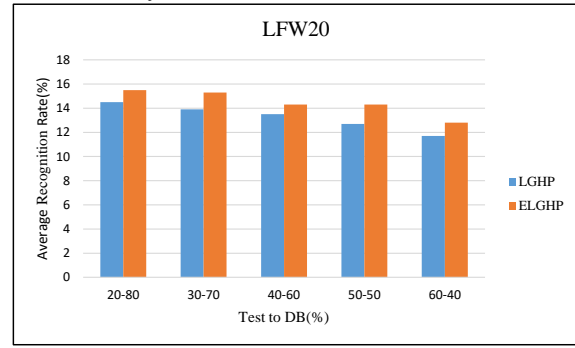

Fig. 10. Comparative average recognition rates of LGHP and ELGHP of 10-fold cross validation with different size test and DB on LFW20.

### A2. Recognition Accuracy on color-FERET database

FERET or Facial Recognition Technology Database which was sponsored by the Department of Defense's Counterdrug Technology Development Program through the Defense Advanced Research Products Agency (DARPA) contains 14051 images with different facial expressions and pose. Similar to LFW, image classes that have at least 20 images belonging to it were take resulting in a total of 4051 images. ELGHP working on 20-80 test database split, with 791 images in test and 3260 images in the database gives an average accuracy of 82.4% with R = 3 and 20% test size compared to LGHP which gives 76.9% accuracy. Fig. 11 shows that improved performance also exists at others splits.

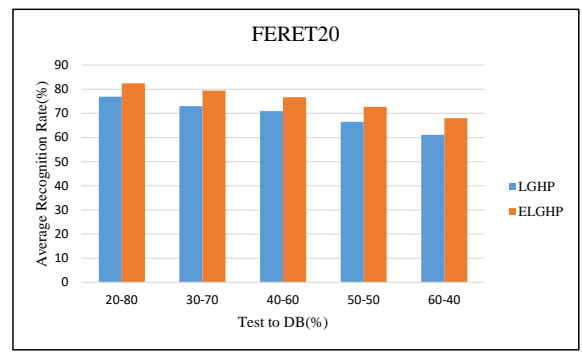

Fig. 11. Comparative average recognition rates of LGHP and ELGHP of 10-fold cross validation with different size test and DB on FERET20.

### A4. Recognition Accuracy on Extended Yale B database

The extended Yale Face Database B contains 16128 images of 28 human subjects under 9 poses and 64 illumination conditions. The performance of both ELGHP and LGHP was recorded on the cropped version of extended Yale B Face Database. On this database ELGHP has 75% accuracy with R = 3 and 20% test size compared to LGHP which recorded 70.4% accuracy. As demonstrated in Fig. 13, ELGHP maintains superiority over LGHP on other splits as well.

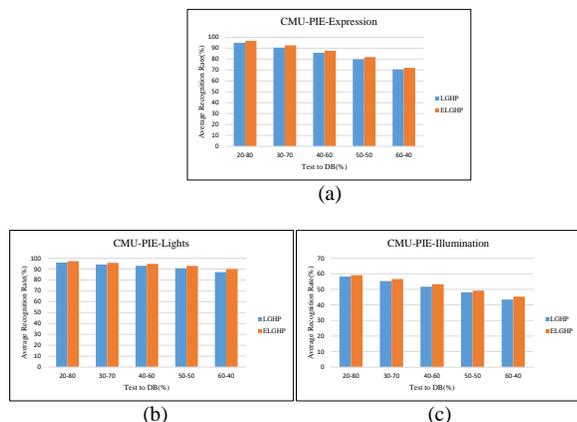

(a)

(b) (c)

Fig. 12. Comparative average recognition rates of LGHP and ELGHP of 10-fold cross validation with different size test and DB on (a) CMU-PIE-Expression, (b) CMU-PIE-Lights, and (c) CMU-PIE-Illumination.

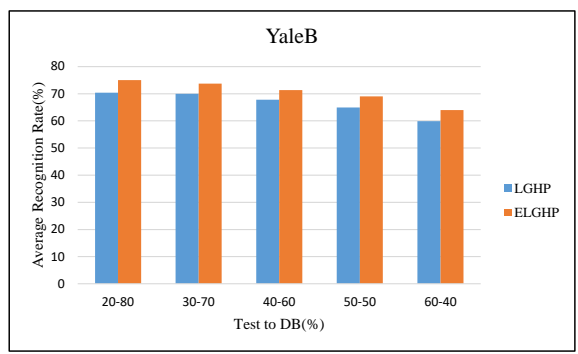

Fig. 13. Comparative average recognition rates of LGHP and ELGHP of 10-fold cross validation with different size test and DB on Extended Yale B

### A5. Recognition Accuracy on Gallagher database

Gallagher Collection Person Dataset contains 931 images of 32 people captured in real life with real expressions. For this experiment, we have taken those individuals who have five or more images in the dataset resulting in a total of 896 images. Due to images captured in real life with a natural varying background, descriptors on Gallagher report very low accuracies. ELGHP has an average accuracy of 63.6% with R = 3 and 20% test size whereas LGHP has 62.3% accuracy. Fig. 14 illustrates similar trend on other splits. It is important to note that ELGHP shows improvement over LGHP in all the five test and gallery splits viz. 20-80, 30-70, 40-60, 50-50 and 60-40 on all the world class databases that we have considered.

### B. Execution Time

To analyze the performance result of implementation level parallelism of ELGHP, we measured the time taken for test feature extraction and classification time as database features in a face recognition system are generally considered to be available at all times for comparison. The test was conducted on 3 databases, LFW, FERET, and CMU-PIE. We ran our C++ implementation with different thread counts allocating half of the threads to test feature extraction and the rest half to classification.

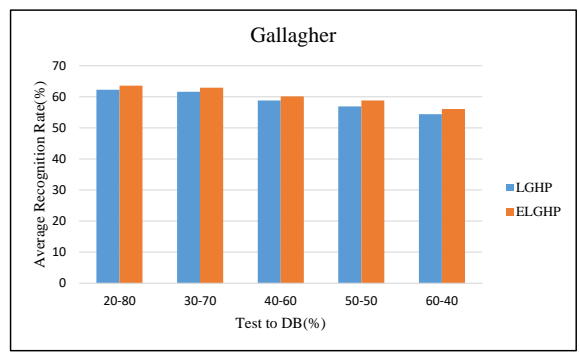

Fig. 14. Comparative average recognition rates of LGHP and ELGHP of 10-fold cross validation with different size test and DB on Gallagher.

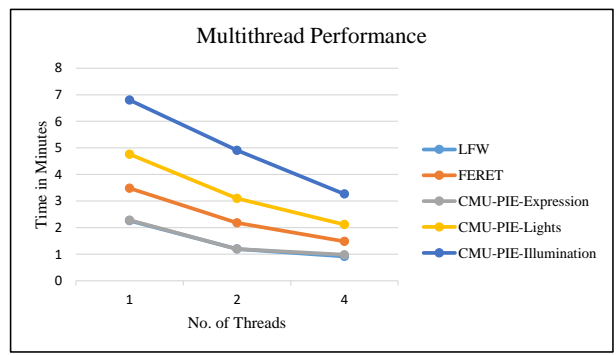

Fig. 15. Single thread vs Multithread execution time comparison of ELGHP on various world class face databases. The execution time of CMU-PIE-Illumination has been reduced to 10% for better representation.

For one thread LFW took 2.267 minutes, FERRET 3.483 minutes, CMU-Pie-Expression: 2.283 minutes, CMU-Pie-Illumination: 68.033 minutes and CMU-Pie Lights: 4.767 minutes. For two threads LFW took 1.2 minutes, FERRET takes 2.183 minutes, CMU-Pie Expression around 1.2 minutes, CMU-Pie-Illumination: 49.133 minutes and CMU-Pie Lights: 3.1 minutes. The execution-time improvement achieved for one thread vs two thread was around 38%. For four threads LFW took 0.917 minutes, FERRET takes 1.483 minutes, CMU-Pie Expression around 0.983 minutes, CMU-Pie Illumination: 32.67 minutes and CMU-Pie Lights: 2.12 minutes. The execution-time improvement achieved for one thread vs four thread was around 60%. As shown in Fig. 15, this result helps to state that parallel processing allows ELGHP to have significant execution time improvement over serial processing. Similar performances were achieved when different threads were applied on different radius on the same pixel. This in particular shows that ELGHP is well suited for parallel processing.

## VI. CONCLUSION

This paper proposes the design of a novel facial biometric system, illustrates the steps required to build such a system, and describes in detail the steps involved in its working and usage. The system incorporates a novel local face descriptor as an extended version of LGHP. Performance accuracy has been analyzed on several world class databases. Implementation level parallelism of extended LGHP has been proposed and performance measured with different thread counts and databases. The system proposed in its current state is limited in its usage and cannot be used for surveillance or multi-face recognition. The system also does nothing to guard against spoofing i.e. distinguish between 'live' face and 'not live' face. Future work on the proposed system will resolve these issues and expand its usage scenarios.

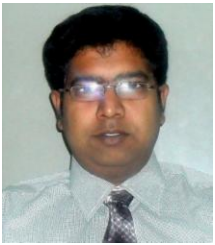

**Soumendu Chakraborty** received his BE degree in Information Technology from University Institute of Technology, University of Burdwan, India. He received his M.Tech. in Computer Science and Engineering from GLA University, Mathura, India. He has 8 years of teaching and 4 years of research experience and currently he is pursuing his PhD from Indian Institute of Information Technology, Allahabad, U.P., India. His research interests include image processing, biometric systems and pattern recognition.

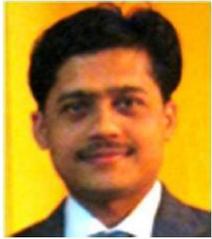

**Satish Kumar Singh** is currently an Assistant Professor with the Indian Institute of Information Technology, Allahabad, India. He received the Ph.D., M.Tech., and B.Tech. degrees in 2010, 2005, and 2003, respectively. He has over 13 years of experience in academic and research institutions. He has authored over 30 publications in reputed international journal and conference proceedings. He is a member of various professional societies, like the Institution of Electronics and Telecommunication Engineers. He was an Executive Committee Member of the 2014 IEEE Uttar Pradesh Section. He is serving as an Editorial Board Member and reviewer for many international journals. His current research interests are in the areas of digital image processing, pattern recognition, multimedia data indexing and retrieval, watermarking, Deep learning and biometrics.

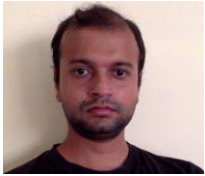

**Kush Kumar** received his B.Tech. degree from NIT Patna, India. He is currently pursuing his M.Tech. in Human Computer Interaction from Indian Institute of Information Technology, Allahabad, U.P., India. Previously he worked as software engineer at Samsung Research, Bangalore, India. His research interests include design patterns, machine learning and image processing.